\documentclass[twocolumn,           
               showpacs,            
               nopreprintnumbers,     
               aps,                 
               prd,          	    
               letterpaper,             
               superscriptaddress,      
               nofootinbib,         
               tightenlines,        
               floats,floatfix      
               ]{revtex4-1}

\usepackage{graphicx}
\usepackage{dcolumn}
\usepackage{bm}
\usepackage{amsmath}
\usepackage{color}
\usepackage{enumerate}

\begin{document}

\title{Stellar stability in brane-worlds revisited}

\author{Miguel A. Garc\'{\i}a-Aspeitia} 
\email{aspeitia@fisica.uaz.edu.mx}
\affiliation{Departamento de F\'isica, DCI, Campus Le\'on, Universidad de 
Guanajuato, C.P. 37150, Le\'on, Guanajuato, M\'exico.}
\affiliation{Consejo Nacional de Ciencia y Tecnolog\'ia, Av, Insurgentes Sur 1582. Colonia Cr\'edito Constructor, Del. Benito Ju\'arez C.P. 03940, M\'exico D.F. M\'exico.}
\affiliation{Unidad Acad\'emica de F\'isica, Universidad Aut\'onoma de Zacatecas, Calzada Solidaridad esquina con Paseo a la Bufa S/N C.P. 98060, Zacatecas, M\'exico.}

\author{L. Arturo Ure\~na-L\'opez}
\email{lurena@fisica.ugto.mx}
\altaffiliation[Present address: ]{Institute for Astronomy, University of Edinburgh, Royal Observatory, Edinburgh EH9 3HJ, United Kingdom.}
\affiliation{Departamento de F\'isica, DCI, Campus Le\'on, Universidad de 
Guanajuato, C.P. 37150, Le\'on, Guanajuato, M\'exico.}

\begin{abstract}
We consider here the general conditions for the stability of brane stars that obey a so called a \emph{minimal setup}: the nonlocal anisotropic stress and energy flux are everywhere absent, and the only permitted Weyl correction is the interior solution of the nonlocal energy density. Along with a series of simple conditions, we show that the Germani-Maartens solution with a constant density sets up the upper bound for the compactness of that particular class of brane stars. The general demonstration is based upon the properties of the interior solutions of the stars, although we also show that the minimal setup implies a Schwarzschild exterior.
\end{abstract}

\draft
\date{\today}
\maketitle

\section{Introduction.} \label{Int}
Brane world dynamics is one of the most interesting possibilities to generalize Einstein's General Relativity (GR), by means of adding in the four dimensional Einstein equations new terms that come from the presence of extra spatial dimensions. The simplest realization is based on the idea that our universe is a five dimensional Anti de Sitter-Schwarszchild (AdS-S) manifold, with one (or more) four-dimensional embedded
branes\cite{Gogberashvili:1998vx,Randall-I,Randall-II,PhysRevLett.84.586,*Binetruy:1999ut,*PhysRevD.62.024011,*Anupam} (for a review see\cite{Maartens,*PerezLorenzana:2004na}). For example, Randall-Sundrum (RS) models\cite{Randall-I,*Randall-II} were intended to alleviate the hierarchy problem between gravitation and the other forces, through the introduction of a warped extra dimension. For instance, the physical mass of a fundamental scalar field (like the Higgs boson), and its vacuum, would naturally appear at TeV scales rather than the Planck scale, without the need of any large hierarchy on the radius\cite{Randall-I,PerezLorenzana:2004na}.

In general terms, the brane additions in the equations of motion include quadratic corrections to the energy-momentum tensor, and non-local effects obtained from the traceless Weyl tensor $\xi_{\mu\nu}$, once the five-dimensional dynamics is projected onto our four-dimensional brane. The new terms can help to alleviate different cosmological puzzles, such as those of dark energy and dark matter\cite{PhysRevLett.90.241301,Okada:2004nc,aspeitia1,*aspeitia2,JuanLuis,*JuanLuis2,Lue,Binetruy,Wang,Schwindt:2005fm,Chakraborty:2007ad}, but they also introduce changes in the stability and dynamics of stellar objects\cite{Germani,Linares,*Aspeitia3,*Castro:2014xza}. Some of these studies show constraints of the brane tension $\lambda$ coming from observations on stellar
equilibrium of about $\lambda \sim 5 \times10^{8} \, {\rm MeV^{4}}$\cite{Maartens,Germani,Aspeitia3}, see also the constraints from classical tests\cite{Bohmer:2009yx,*Boehmer:2008zh}. Cosmological studies from nucleosynthesis provide the lower bound $\lambda > 1 \, {\rm MeV^{4}}$\cite{Maartens,Anupam}. Both results are complementary and point out to the energy region in which brane dynamics could be possibly observed.

In this paper we revisit the stability conditions for star that obeys the brane-modified Einstein's equation, in a calculation that parallels that of textbooks for the case of GR and without the need to deal directly with the modified Tolman-Oppenheimer-Volkoff equation (see for example\cite{Castro:2014xza}). We shall write a \emph{minimal setup} of simple but physically motivated assumption that allow us to establish the existence of an interior solution, which is then used combination with the equations of motion to write a constraint equation for the compactness of the brane stars.

\section{Stellar Stability}
Let us start by writing the equations of motion for stellar stability in a brane embedded in a five-dimensional bulk according to the Randall-Sundrum II model\cite{Randall-II}. Following an appropriate computation (for details see\cite{Maartens,*Shiromizu}), it is possible to demonstrate that the modified four-dimensional Einstein's equations can be written as 
\begin{equation}
  G_{\mu\nu} + \Lambda_{(4)}g_{\mu\nu} = \kappa^{2}_{(4)} T_{\mu\nu} + \kappa^{4}_{(5)} \Pi_{\mu\nu} +
  \kappa^{2}_{(5)} F_{\mu\nu} - \xi_{\mu\nu} , \label{Eins}
\end{equation}
where $T_{\mu \nu}$ is the four-dimensional energy-momentum tensor of the matter trapped in the brane, and the rest of the terms on the right-hand-side are explicitly given by:
\begin{subequations}
\label{1}
\begin{eqnarray}
\Lambda_{(4)}&=&\frac{1}{2}\Lambda_{(5)}+\frac{\kappa^{4}_{(5)}}{12}\lambda^{2}, \label{1a} \\
\kappa^{2}_{(4)}&=&8\pi G_{N}=\frac{\kappa^{4}_{(5)}}{6}\lambda, \label{1b} \\
\Pi_{\mu\nu}&=&-\frac{1}{4}T_{\mu\alpha}T^{\alpha}_{\nu}+\frac{TT_{\mu\nu}}{12}+\frac{g_{\mu\nu}}{24}(3T_{\alpha\beta}T^{\alpha\beta}-T^{2}), \label{1c} \\
F_{\mu\nu}&=&\frac{2T_{AB}g_{\mu}^{A}g_{\nu}^{B}}{3}+\frac{2g_{\mu\nu}}{3}\left(T_{AB}n^{A}n^{B}-\frac{^{(5)T}}{4}\right), \label{1d} \\
\xi_{\mu\nu}&=&^{(5)}C_{AFB}^{E}n_{E}n^{F}g_{\mu}^{A}g_{\mu}^{B}. \label{1e}
\end{eqnarray}
\end{subequations}

One one hand, we see in Eq.~\eqref{1a} the typical relationship among $\Lambda_{(4)}$, $\Lambda_{(5)}$, which are the five and four dimensional cosmological constants, respectively, and $\lambda$, which is related to the brane tension. Likewise, Eq.~\eqref{1b} shows the relationship between $\kappa_{(4)}$ and $\kappa_{(5)}$, which are respectively the four and five-dimensional coupling constants of gravity, where $G_{N}$ is Newton's gravitational constant.

On the other hand, $\Pi_{\mu\nu}$ represents the quadratic corrections on the brane generated from the four-dimensional energy-momentum tensor $T_{\mu \nu}$, whereas $F_{\mu\nu}$ gives the contributions of the energy-momentum tensor in the bulk $T_{AB}$ (with latin letters taking values $0,1,2,3,4$), which is then projected onto the brane with the help of the unit normal vector $n_{A}$. Finally, $\xi_{\mu\nu}$ gives the contributions of the five-dimensional Weyl's tensor $^{(5)}C^{E}_{AFB}$ when also projected onto the brane manifold (see\cite{Shiromizu} for more details).

For purposes of simplicity, we will not consider bulk matter and then $T_{AB} = 0$, which translates into $F_{\mu\nu}=0$, and will also discard the presence of the four-dimensional cosmological constant, $\Lambda_{(4)} = 0$, as we do not expect it to have any important effect at astrophysical scales (for a recent discussion about this see\cite{Pavlidou:2013zha}). Additionally, we will neglect any nonlocal energy flux, which is allowed by the spherical symmetry of the solutions we will study below\cite{Germani}.

The energy-momentum tensor $T_{\mu \nu}$, the quadratic energy-momentum tensor $\Pi_{\mu \nu}$, and the Weyl (traceless) contribution $\xi_{\mu\nu}$, have the explicit forms
\begin{subequations}
\label{eq:4}
\begin{eqnarray}
\label{Tmunu}
T_{\mu\nu} &=& \rho u_{\mu}u_{\nu} + p h_{\mu\nu} \, , \\
\label{Pimunu}
\Pi_{\mu\nu} &=& \frac{1}{12} \rho \left[ \rho u_{\mu}u_{\nu} + (\rho+2p) h_{\mu\nu} \right] \, , \\
\label{ximunu}
\xi_{\mu\nu} &=& - \frac{\kappa^4_{(5)}}{\kappa^4_{(4)}} \left[ \mathcal{U} u_{\mu}u_{\nu} + \mathcal{P}r_{\mu}r_{\nu}+ \frac{ h_{\mu\nu} }{3} (\mathcal{U}-\mathcal{P} ) \right] \, .
\end{eqnarray}
\end{subequations}
Here, $p$ and $\rho$ are, respectively, the pressure and energy density of the stellar matter of interest, $\mathcal{U}$ is the nonlocal energy density, and $\mathcal{P}$ is the nonlocal anisotropic stress. Also, $u_{\alpha}$ is the four-velocity (that also satisfies the condition $g_{\mu\nu}u^{\mu}u^{\nu}=-1$), $r_{\mu}$ is a unit radial vector, and $h_{\mu\nu} = g_{\mu\nu} + u_{\mu} u_{\nu}$ is the projection operator orthogonal to $u_{\mu}$.

Spherical symmetry indicates that the metric can be written as:
\begin{equation}
{ds}^{2}= - B(r){dt}^{2} + A(r){dr}^{2} + r^{2} (d\theta^{2} + \sin^{2} \theta d\varphi^{2}) \, .\label{metric}
\end{equation}
If we define the reduced Weyl functions $\mathcal{V} = 6 \mathcal{U}/\kappa^4_{(4)}$, and $\mathcal{N} = 4 \mathcal{P}/\kappa^4_{(4)}$, then the equations of motion for a relativistic star in the brane are:
\begin{subequations}
  \label{eq:7}
\begin{eqnarray}
  \mathcal{M}^\prime &=& 4\pi{r}^{2}\rho_{eff} \, \label{eq:7d} \, , \\
  \frac{B^{\prime}}{B} &=& \frac{2G_N}{r^{2}} \frac{4 \pi \, p_{eff} \, r^3 + \mathcal{M}}{1 - 2G_N \mathcal{M}/r} \, ,   \label{eq:7a} \\
  p^\prime &=& -\frac{1}{2} \frac{B^{\prime}}{B} ( p + \rho ) \, ,   \label{eq:7b} \\
  \mathcal{V}^{\prime} + 3 \mathcal{N}^{\prime} &=& - \frac{B^{\prime}}{B} \left( 2 \mathcal{V} + 3 \mathcal{N} \right) - \frac{9}{r} \mathcal{N} - 3 (\rho+p) \rho^{\prime} \, ,  \label{eq:7c}
\end{eqnarray}
\end{subequations}
where a prime indicates derivative with respect to $r$, $A(r) = [1 - 2G_N \mathcal{M}(r)/r]^{-1}$, and the effective energy density and pressure, respectively, are given as:
\begin{subequations}
\label{eq:3}
\begin{eqnarray}
\rho_{eff}  &=& \rho \left( 1 + \frac{\rho}{2\lambda} \right) + \frac{\mathcal{V}}{\lambda}  \, , \label{eq:3a} \\
p_{eff} &=& p \left(1 + \frac{\rho}{\lambda} \right) + \frac{\rho^{2}}{2\lambda} + \frac{\mathcal{V}}{3\lambda} + \frac{\mathcal{N}}{\lambda} \, . \label{eq:3b}
\end{eqnarray}
\end{subequations}

To demonstrate the general conditions for stellar stability, we follow closely the standard textbook procedure that has been established within GR, see for instance\cite{weinberg:1972}. The main physical requirements are:
\begin{enumerate}[(a)]
\item The radius $R$ is fixed, with $\rho(r) = 0$ for $r > R$.

\item The pressure vanishes at the surface and in the exterior of the star, and then $p(r) =0$ for $r \geq R$.

\item The total mass $M = \mathcal{M}(R)$ is fixed.

\item The metric coefficient $A(r)$ must not be singular, which implies $\mathcal{M}(r)<r/2G_N$.

\item The energy density $\rho(r)$ must not increase outwards:  $\rho^{\prime} (r) \leq 0$.

\item A null nonlocal anisotropic stress in the interior of the star: $\mathcal{N}(r) =0$ for $r \leq R$.

\item The non-local energy density vanishes at the surface, and then $\mathcal{V}(R) = 0$.
\end{enumerate}

We will follow conventional wisdom and consider both conditions (a) and (b) to be physically reasonable assumptions. However, other possibilities might have been chosen. For instance, in the case of condition (a) we could have considered instead that $\rho_{eff}(r) = 0$ for $r > R$, but this is a more restrictive assumption as it would have implied that $\mathcal{U}(r) =0$ for $r> R$, see Eq.~\eqref{eq:3a}. Thus, condition (a) allows us to consider implicitly a wider set of solutions in our demonstration for stellar stability.

The case is not that clear for condition (b), as $p(R) =0$ is not a necessary condition in brane stars as is in GR. In fact, the Israel-Darmois matching condition at the surface requires the continuity of the effective pressure: $p_{eff}(R^-) = p_{eff}(R^+)$\cite{Germani}, and then prior knowledge about the Weyl terms would be needed for a precise determination of $p(R)$, see Eq.~\eqref{eq:3b}. Nonetheless, we believe condition (b) is physically reasonable and not excessively restrictive, and then we will simply enforce it upon our solutions.

Conditions (c) and (d) refers to the total mass of the stellar configuration that includes the brane corrections (quadratic plus Weyl terms), and not to the physical mass that arises from the direct integration of the density $\rho(r)$. That is more convenient in the general case in which one cannot make an explicit separation of the brane corrections, and, in addition, provides a clear physical interpretation of the total mass in those brane stars that have an exterior Schwarzschild solution. Condition (e) is the expected requirement for the physical energy density in a stellar configuration, as is a necessary condition in our brane solutions to obtain a finite total mass, see Eqs.~\eqref{eq:7d} and~\eqref{eq:3a}.


Condition (f) is just chosen for convenience and simplicity in the calculations; however, the nonlocal anisotropic stress $\mathcal{N}$ is for now allowed to have an exterior non-trivial solution. Finally, condition (g) is the simplest possibility for the boundary condition of the nonlocal energy density $\mathcal{V}$, and from it Eq.~\eqref{eq:7c} generates the solution:
\begin{equation}
  \label{eq:9}
    \mathcal{V} (r) = \frac{3}{B^2(r)} \int^R_r B^2 (\rho + p) \rho^\prime \, dr \, ,
\end{equation}
for $r < R$. Condition (e) implies that $\mathcal{V} (r <R) \leq 0$, and, from Eq.~\eqref{eq:7c}, also that $\mathcal{V}^\prime (r <R) \geq 0$. It can be seen that conditions (a)-(g) are very general premises to solve the equations of motion~\eqref{eq:7}, and we consider them to be the \emph{minimal setup} for stars with the least of brane corrections (for other possibilities see\cite{Wiseman:2001xt,*Ovalle:2013vna,*HeydariFard:2009tj,*Deruelle:2001fb}).

In the standard case of GR, it can be shown that there is a unique solution of the equations of motion for a given function $\rho(r)$. Then, any general bound on the compactness of the star can be determined by choosing an appropriate $\rho(r)$ to provide the most extreme case, and this results in the solution with a constant density. It is important to emphasize that compactness refers here to the minimum radius $R$ that a stellar configuration can have for a given total mass $M$, and which can be quantified in GR by the upper bound $G_{N}M/R<4/9$\cite{weinberg:1972}.

In the brane case depicted here, the same argument applies as the behavior of functions $p(r)$, $B(r)$, and $\mathcal{V}(r)$, can be directly determined from conditions (a)-(g) once we choose a given $\rho(r)$. To begin with, condition (b) fixes the boundary condition of the pressure at the surface of the star, and likewise condition (c) implies that $A^{-1}(R) = 1 - 2G_N M/R$; then the value of $B(R)$ can be tuned up appropriately to finally integrate Eqs.~\eqref{eq:7a} and~\eqref{eq:9}. Conditions (e)-(g) then establish that a full interior solution of the equations of motion~\eqref{eq:3}, with boundary conditions at the surface, can be uniquely found once we choose a particular form of $\rho(r)$. 

We are now in a position to start the general demonstration for stability. If we take $B \equiv \zeta^{2}$, then Eq.~\eqref{eq:7a} can be written as
\begin{equation}
  \label{eq:12}
\frac{d}{dr} \left[ \frac{1}{A^{1/2}(r)r} \frac{d\zeta(r)}{dr} \right] = G_N A^{1/2}(r) \left( \frac{\mathcal{M}(r)}{r^{3}} \right)^{\prime} \zeta(r) \, .
\end{equation}
From here, we proceed to derive an upper bound for $\zeta(0)$. If $\zeta$ is positive, then the right-hand side of Eq. \eqref{eq:12} is negative, because the mean effective density cannot increase with $r$ if the effective density does not. This can be seen from the fact that, even if $\mathcal{V}^\prime (r<R) \geq 0$, the latter is mediated by the brane tension and then we expect that $\rho^\prime_{eff} (r<R) \leq 0$ will follow from condition (e). Hence, we obtain the condition
\begin{equation}
\frac{d}{dr} \left[ \frac{1}{A^{1/2}(r)r} \frac{d\zeta(r)}{dr} \right] \leq 0 \, . \label{may1}
\end{equation}
To solve Eq. \eqref{may1}, we need the explicit value of $\zeta^\prime(R)$, which is obtained from Eq. \eqref{eq:7a} as
\begin{equation}
  \label{eq:2}
  \zeta^\prime (R) = \zeta(R) \frac{G_N}{R^{2}} \left[ \frac{ M + 4\pi \, p_{eff}(R^-) R^{3}}{1 - 2G_N M/R} \right] \, .
\end{equation}
The integration of Eq. \eqref{may1} from $r=0$ to $r=R$ yields
\begin{eqnarray}
&&\frac{\zeta(0)}{\zeta(R)}  \leq  1 - \frac{G_N}{R^{3}} \times \nonumber\\
 && \left[ \frac{ M + 4\pi \, p_{eff}(R) R^{3}}{\sqrt{1 - 2G_N M/R}} \right] \int_{0}^{R} \frac{r \, dr}{\sqrt{1 - 2G_N \mathcal{M}(r)/r}} \, . \label{Final}
\end{eqnarray}
Notice that Eq.~\eqref{Final} is a general result that looks very similar to the one in GR, except for the value of the effective pressure at the surface of the star, which we expect to be non null in general. In fact, from our conditions (b), (f), and (g), we find that $p_{eff}(R) = \rho^2(R)/2\lambda$, see Eq.~\eqref{eq:3b}.

As in the textbook case, we must now search for the smallest possible function $\mathcal{M}(r)$ that gives the largest result in the integral on the right-hand side of Eq.~\eqref{Final}, which will in turn provide the smallest upper bound for $\zeta(0)$. For a given configuration with mass $M$, radius $R$, and a density distribution that satisfies $\rho^{\prime}_{eff}(r) \leq 0$, the mass function $\mathcal{M}(r)$ that would be everywhere as small as possible is of the form $\mathcal{M}(r)= M r^{3}/R^{3}$. Then, condition that $\zeta(r) \geq 0$ transforms Eq.~\eqref{Final} into
\begin{equation}
  \label{eq:14}
   \frac{G_N M}{R} < \frac{1}{2} - \frac{1}{18} \left( \frac{1 + 3 p_{eff}(R)/\bar{\rho}_{eff}}{1 + p_{eff}(R)/\bar{\rho}_{eff}} \right)^2 \, ,
\end{equation}
where we have defined the mean energy density as $\bar{\rho}_{eff} \equiv 3M/(4\pi R^3)$. The standard result of GR is obtained from $p_{eff}(R) =0$, and then $G_N M/R < 4/9$.  Eq.~\eqref{eq:14} aroused from the assumption of constant effective density, $\rho_{eff} = \mathcal{M}(r)/r^3 = \mathrm{const.}$, which can be obtained in the extreme case of constant $\rho$ and $\mathcal{V}(r<R) =0$ (see Eq.~\eqref{eq:9}). This in turn implies that $\bar{\rho}_{eff} = \rho (1 + \rho/2\lambda)$, and $p_{eff}(R) = \rho^{2}/2\lambda$, see Eqs.~\eqref{eq:3}, and then
 \begin{equation}
  \frac{G_N M}{R} < \frac{1}{2} - \frac{1}{18} \left( \frac{1 + 2 \rho/\lambda}{1 + \rho / \lambda} \right)^2 \, . \label{finsol}
\end{equation}

Eq.~\eqref{finsol} is exactly the same upper bound on the compactness obtained in the Germani-Maartens (GM) interior solution in Ref.\cite{Germani}, which arises from the assumption of constant density and null Weyl corrections (i.e. $\mathcal{V}(r) \equiv 0 \equiv \mathcal{N}(r)$) in the brane equations of motion. Similarly to the GR case, Eq.~\eqref{eq:14} shows in addition that the GM solution, which is the simplest interior solution with constant density, provides the most general upper bound on the compactness of any brane star that accomplishes the above conditions (a)-(g). 

\begin{figure}[htp]
\centering
\includegraphics[scale=1.0]{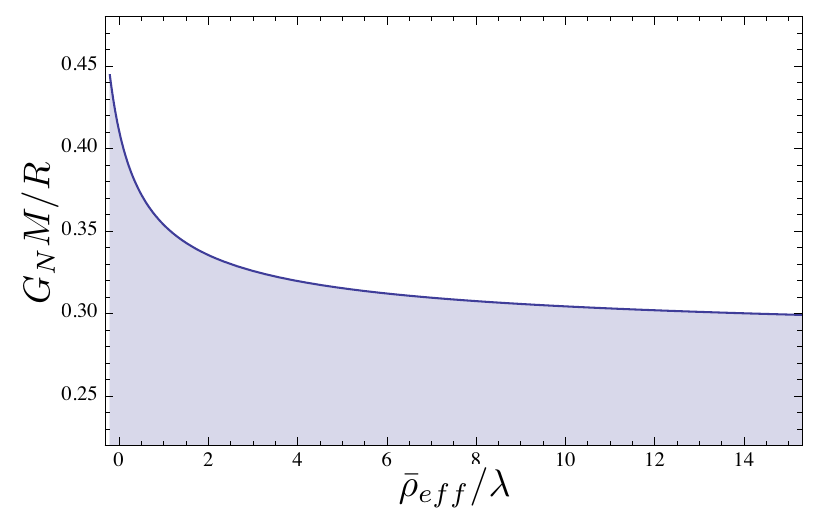}
\caption{Behavior of Eq. \eqref{eq:15} for the gravitational potential $G_{N}M/R$, the shaded area refers to the region where the star can be gravitationally stable. As expected, the GR bound $G_{N}M/R < 4/9$ is obtained in the limit $\bar{\rho}_{eff}/\lambda \to 0$, but the opposite limit  $\bar{\rho}_{eff}/\lambda \to \infty$ gives $G_{N}M/R < 5/18$. Then, a very low value of the brane tension would imply a lower bound on stellar stability.}\label{fig1}
\end{figure}

A more useful expression, that can have a wider applicability to other star solutions, can be given in terms of the mean energy density:
\begin{equation}
  \label{eq:15}
  \frac{G_N M}{R} < \frac{1}{2} - \frac{1}{18} \left( \frac{4 \sqrt{ 1 + 2\bar{\rho}_{eff}/\lambda} -3}{2 \sqrt{ 1 + 2\bar{\rho}_{eff}/\lambda} -1} \right)^2 \, ,
\end{equation}
which is illustrated in Fig.~\ref{fig1}. It must be stressed out that the mean energy density $\bar{\rho}_{eff}$, is defined solely from the total radius $R$ and total mass $M$ of the star, where the latter includes the brane effects carried out by the interior solution of the Weyl function $\mathcal{V}$, see Eq.~\eqref{eq:9}.
 
Notice that Eq.~\eqref{eq:15} typically shows that the upper bound on the compactness of a star is not an absolute number as in GR, but that we need information about the effective density of a brane star before assessing its stability. Eq.~\eqref{eq:15} allows a very compact star to have an effective density either larger or smaller than the brane tension, and then we cannot use it to put constraints on the latter unless we get astrophysical information on stars with a compactness in the range $ 5/18 < G_NM/R < 4/9$. Even a typical neutron star, with parameters $\bar{\rho}_{eff} \sim 10^{9} \, \textrm{MeV}^4$ and $M \sim 10^{57} \, \textrm{GeV}$ (for recent examples of the detection of neutron stars see\cite{Guver:2013xa,*Ozel:2011ht,*Guver:2010td,*Guillot:2013wu,*Steiner:2012xt}), has a compactness of about $G_N M/R \sim 0.1 < 5/18$, and then it cannot be used to give sensitive information about astrophysical limits on $\lambda$.

To get Eq.~\eqref{eq:15} we only needed to know the main properties of the interior solutions that can be obtained from Eqs.~\eqref{eq:7}. However, the exterior functions can be described in general as follows. To start with, we write again the Israel-Darmois matching condition, which under conditions (a), (b), (f), and (g), translates into
\begin{equation}
  \label{eq:1}
  \mathcal{V}(R^+) + 3 \mathcal{N}(R^+) = (3/2) \rho^2(R) \, .
\end{equation}
As shown in\cite{Germani}, the case of constant density requires an exterior solution with non-trivial expressions for the Weyl functions, and then the GM solution cannot be matched up with a Schwarzschild exterior. Actually, there exist for the GM solution at least two exterior solutions with both Weyl functions different from zero.

However, the case is different for realistic stars with both a non-constant density and $\rho(R) =0$, as for them the Israel-Darmois matching condition states that $\mathcal{V}(R^+) + 3 \mathcal{N}(R^+) = 0$, which is not but the requirement that the effective pressure, following the physical pressure, vanishes at the surface of the star: $p_{eff}(R)=0$. The simplest exterior solution of Eq.~\eqref{eq:7c} that satisfies this boundary condition is the trivial one: $\mathcal{V}(r) = 0 = \mathcal{N}(r)$ for $r > R$. In other words, the exterior of realistic brane stars can just be the typical Schwarzschild solution $B(r) = A^{-1}(r) = 1- 2G_N M/r$, and then the total mass $M$ to be used in Eq.~\eqref{eq:15} could be determined, for instance, from the orbits of test particles around the star. 

As a final note, it can be seen that our consideration of a minimal setup eventually implied the total exclusion of the nonlocal anisotropic stress $\mathcal{N}(r)$ from the full brane solution of realistic stars, mainly because of the interplay between condition (f) and the Israel-Darmois matching constraint.

Nonetheless, the question remains of whether we could have a second version of the minimal setup under the following new conditions: (f) A null nonlocal energy term in the interior of the star: $\mathcal{V}(r) =0$ for $r \leq R$, and (g) The nonlocal anisotropic stress vanishes at the surface: $\mathcal{N}(R) = 0$. The last condition is in agreement with the Israel-Darmois matching condition in the case of realistic stars, but this time Eq.~\eqref{eq:7c} also requires: (h) $\mathcal{N}(0) =0$, in order to have a non-divergent solution at the center of the star. Indeed, the regular solution of Eq.~\eqref{eq:7c} under the above condition (h) can be written in quadratures as:
\begin{equation}
\label{N}
\mathcal{N}(r)= \frac{1}{B(r)r^{3}} \int_0^{r} B r^{3} (\rho+p) \rho^{\prime} dr \, ,
\end{equation}
and then it is not possible to accomplish at the same time the aforementioned condition $\mathcal{N}(R) = 0$. This is an inescapable and problematic situation, as the equation of motion~\eqref{eq:7c} is a first order differential equation that is now tied up by two boundary conditions, and it is not possible to guarantee the existence of a solution in the general case that is also able to satisfy both of them. It seems then that the original conditions (a)-(g) established above conform the only \emph{minimal setup} of brane corrections for realistic stars.

\section{Conclusions}
The above discussion opens interesting points about the calculations shown here. First of all, although the GM solution shares with realistic brane stars all the conditions (a)-(g), and was fundamental to establish the general upper bound on the compactness of brane stars, it must be realized that it does not really belong to the class of realistic brane stars that it helped to bound up. Second, we have argued that the exterior of realistic brane stars must be Schwarzschild, as this is the simplest possibility allowed by the equations of motion and the boundary conditions. If this were the case, then classical gravitational tests based upon the exterior solution would be unable to constraint brane effects in realistic stars, except for those implicitly involved in the mass term $M$. In particular, the use of the black hole solution found in\cite{Dadhich:2000am} would not be the most appropriate approach to constraint astrophysical brane effects\cite{Bohmer:2009yx,Boehmer:2008zh}. 

As we also mentioned in the introduction, former studies considered the case of stars with constant density, in which all brane contributions can be singled out so that the brane tension seems to be bounded from below as $\lambda \sim 5 \times10^{8} {\rm MeV^{4}}$\cite{Germani,Aspeitia3}. This procedure cannot be done for realistic stars, for which Eq.~\eqref{eq:15} appears as a more appropriate constraint for their compactness.

The revisited analysis presented in this paper opens a new window to find the bound of stellar stability in brane-world theories, and provides a more general framework that includes topological terms. It also extends the GR case for a particular class of brane stars, and shows the usability of the GM solution as the plausible extreme case to constraint their compactness. The procedure can be replicated for other classes of stars that may include the presence of both nonlocal Weyl terms and whose solutions may also require extra numerical analysis. This is ongoing research that will be presented elsewhere.

\begin{acknowledgements}
MAG-A acknowledges support from a CONACyT postdoctoral grant, C\'atedra-CONACYT and SNI. LAU-L thanks the Royal Observatory, Edinburgh, where part of this work was done during a sabbatical stay, for its kind hospitality. This work was partially supported by PROMEP, DAIP, PIFI, and by CONACyT M\'exico under grants 167335 and 179881. We also thank the support of the Fundaci\'on Marcos Moshinsky, and the Instituto Avanzado de Cosmolog\'ia (IAC) and the Beyond Standard Theory Group (BeST) collaborations.
\end{acknowledgements}

\bibliography{librero1}

\end{document}